\documentstyle[twocolumn,aps,prl]{revtex}

\begin{document}
\pagestyle{empty}

\title{
An Anyon Superconducting Groundstate in Si MOSFETS?\hfill\break
F.C.  Zhang$^{1}$ and T.M. Rice$^{2}$\hfill\break}

\address{$^{1}$Department of Physics, University of Cincinnati,
  Cincinnati, OH 45221\\ 
$^{2}$Theoretische Physik, ETH-H\"onggerberg, 8093 Z\"urich,
Switzerland\\
\vspace{.3cm}
}

\maketitle

\narrowtext 

The recent report by  Kravchenko et al.~\cite{Kravchenko} 
of a metal-insulator transition (MIT) in 2D Si-MOSFETS 
with decreasing electron density has caused a lot of interest 
in possible exotic groundstates of a 2D electron gas.  As 
emphasized by Dobrosavljevic
et al.~\cite{Abrahams} the standard scaling analysis for a 2D 
conductor allows only insulating or perfectly conducting  
behavior at T=0K but not a finite conductance.
The data of Kravchenko et al.~\cite{Kravchenko} are consistent with
scaling theory over a wide range of conductance values for electron
densities $n\sim n_c$. The critical density,
$n_c$, however is quite low so that the Coulomb interaction dominates the
kinetic energy of the electrons leading one to search for possible
superconducting (SC)
groundstates stabilized by the Coulomb interactions.  This point of view is
strengthened by the recent report
of Ismail et al.~\cite{Washburn} of a MIT at a similar value of 
$n_c ( ~ 2\times10^{11} cm^{-2})$ in high mobility samples with values of 
$k_Fl \approx 40$ ($k_F$: Fermi wavevector, $l$: mean free path)
at $n =n_c$, much higher than the original value ($k_Fl \approx
1$)~\cite{Kravchenko}. 
In this Comment we wish to draw attention to an earlier proposal
by Ren and Zhang~\cite{Zhang} for a spontaneous flux state (SFS)
closely related to a 
spin-unpolarized $1/2$-filled quantum Hall state (QHS) which would 
be an anyon SC.
Alternative proposals for a $p$-wave SC~\cite{Wang} and for a novel
spin-triplet but even-parity SC~\cite{Belitz} have been put forward
recently. 

Ren and Zhang~\cite{Zhang} showed that this time reversal breaking state could
be stabilized by Coulomb interactions. In its simplest form the ground state 
wavefunction of a SFS can be written as 
$$ \Psi =\left[\prod_{i<j}
  |z_i-z_j|^2/\left(z_i-z_j\right)^2\right]\Psi_S\ ,
$$ 

\noindent
where $\Psi_S$ is a spin unpolarized QHS with the total Landau level (LL)
filling  $\nu=1/2$ (filling 1/4 for each spin). 
Several forms for  $\Psi_S$ have been 
proposed, e.g. by Belkhair and Jain~\cite{Jain},
$\Psi_S=P \psi_1 \psi_2 \psi_{1,1}$, where $\psi_n$ denotes a n-filled LL
of spinless electrons, $\psi_{1,1}$
is the wavefunction with one filled LL for each spin, P is the
projection operator onto the lowest LL. The short range correlations
between electrons is enhanced in a SFS which reduces the interaction
energy.   
Note this state has a gap for single
particle excitations but not for two electron excitations with a total
spin, $S=0$. The form of $\Psi_s$  contrasts with the $\nu=1/2$ spin
polarized QHS,  which is a form of Fermi liquid as discussed 
by Halperin, Lee, and Read~\cite{Halperin}.  

Instead the SFS $\Psi$ is a form of anyon SC, first proposed by
Laughlin~\cite{Laughlin}. Further since it is based on a QHS which is
robust against the presence of impurities, strong sensitivity to
impurities is not expected.

In this scenario the MIT at $n=n_c$ is from a SFS at $n >n_c$ to a
Wigner crystal (or possibly a precursor spin and charge density wave
state) at $n < n_c$. Impurity pinning would be responsible
for the insulating behavior. The higher mobility samples should be in
the weak pinning regime, in which case a depinning transition could
occur with increasing electric field. Note such an interpretation has
been put forward by Pudalov et al.~\cite{Pudalov} to explain their
nonlinear I--V characteristics. 

The report of a MIT in clean samples by Ismail et al.~\cite{Washburn}
at a low
density favors proposals in which the SC originates in the Coulomb
interaction rather than from disorder~\cite{reply}. This is clearly
the case with the scenario described above. The condensation energy of
a SFS arises from a reduction of the Coulomb interaction between
neighboring electrons. 

The question of the behavior at higher densities is not so clear. In
this density regime, the kinetic energy will become dominant and one
might expect a disordered insulator as in single electron theory in
2D. However, Finkel'stein~\cite{Finkelstein} has pointed out that when
one includes 
interactions these scale to strong coupling. Therefore the question of
a second MIT at high density to a disordered insulator is open and
very interesting.

We wish to thank X.C. Xie, M. Ma, K. Ensslin, and A.C. Mota for stimulating
discussions.

\end{document}